\documentclass[aps,pre,reprint,groupedaddress]{revtex4-2}

\usepackage{amsmath}
\usepackage{amssymb}
\usepackage{caption}
\usepackage{subcaption}
\usepackage{graphicx}
\usepackage{xcolor}
\usepackage{CJKutf8}
\usepackage{booktabs}

\begin{document}

\title{Spatial Applications of Topological Data Analysis: Cities, Snowflakes, Random Structures, and Spiders Spinning Under the
Influence}

\author{Michelle Feng}
\affiliation{University of California, Los Angeles}
\author{Mason A. Porter}
\affiliation{University of California, Los Angeles}

\date{\today}


\begin{abstract}

Spatial networks are ubiquitous in social, geographical, physical, and biological applications. To understand the large-scale structure of networks, it is important to develop methods that allow one to directly probe the effects of space on structure and dynamics. Historically, algebraic topology has provided one framework for rigorously and quantitatively describing the global structure of a space, and recent advances in topological data analysis (TDA) have given scholars a new lens for analyzing network data. In this paper, we study a variety of spatial networks --- including both synthetic and natural ones --- using novel topological methods that we recently developed for analyzing spatial networks. We demonstrate that our methods are able to capture meaningful quantities, with specifics that depend on context, in spatial networks and thereby provide useful insights into the structure of those networks, including a novel approach for characterizing them based on their topological structures. We illustrate these ideas with examples of synthetic networks and dynamics on them, street networks in cities, snowflakes, and webs spun by spiders under the influence of various psychotropic substances.

\end{abstract}

\maketitle



\section{Introduction}
\label{sec:intro}

Many complex systems have a natural embedding in a low-dimensional space or are otherwise influenced by space, and it is often insightful to study such spatial complex systems using the formalism of networks \cite{barthelemy2018,newman2018}. In a spatial network, the location of nodes and edges in space can heavily inform both the structure of the network and the behavior of dynamical processes on it. Indeed, obtaining a meaningful understanding of power grids \cite{PhysRevE.77.026102, Kim2018, PhysRevE.69.025103}, granular systems \cite{lia2018}, rabbit warrens \cite{lee2014}, and many other systems is impossible without considering the physical relationships between nodes in a network. For example, to examine traffic patterns on a transportation network in a meaningful way, it is important to include information about the physical distances between points and about the locations and directions of paths between heavily trafficked areas \cite{Batty}.

There are a variety of existing perspectives for studying spatial networks \cite{barthelemy2018,Barthelemy2011}. Many of these perspectives hail from quantitative geography \cite{Haggett1969,pumain2020}. In the 1970s, geographers were already studying the role of space in the formation of networks and in the activities of individuals and goods over geographical networks. As data have become richer and more readily available, it has become possible to take increasingly intricate computational approaches to the study of spatial networks, and a variety of complex-systems approaches have contributed greatly to the literature on spatial networks \cite{barthelemy2018}. Researchers have also proposed various random models for spatial networks, and studying them yields baseline examples to compare to empirical networks \cite{liu2019, nauer2019, community1, community2}. There have also been investigations of the effects of certain spatial network properties on the behaviors of several well-known dynamical processes, including the Ising model \cite{Boulatov1987}, coupled oscillators \cite{Arenas2008}, and random walks \cite{ying2019}.

Although there is much existing work on the properties of spatial networks (e.g., degree distributions, shortest paths, and so on), there are relatively few network tools that leverage ``global'' structure in the traditional topological sense of the word. Current tools for studying global network structure tend to rely on aggregating local information in some way to paint a global picture of a network. By contrast, methods for understanding the global structure of a topological space rely intrinsically on information about the entire space. To illustrate the difference, consider a sphere. If we sample a neighborhood of any point on a sphere, we obtain a surface with the same properties as a plane. If we take a collection of a sphere's neighborhoods (which each resemble a plane) and stitch them together, we are able to obtain a lot of information about the sphere, but we are unable to describe the void in the center of the sphere. (For example, a stereographic projection of a sphere covers the sphere's entire surface, but it fails to capture the void.) To fully understand the structure of a sphere, we must consider the entire sphere at once. Over the last few decades, algebraic topology has been very useful for characterizing the global structure of mathematical spaces \cite{hatcher2002algebraic,edelsbrunner2010} through its use of mathematical tools that consider spaces as global objects. By reframing spatial networks using the language of topological spaces, we can leverage existing topological tools to better understand their structures. For a case study with voting data, see our recent paper \cite{feng2019}.
 
Homology groups, which were defined originally in algebraic topology and have been applied insightfully to a broad range of mathematical topics, provide one way to distinguish between mathematical spaces based on their numbers and types of ``holes'' \cite{hatcher2002algebraic}. Moreover, the extension of homology to so-called ``persistent homology'' (PH) allows one to quantify holes in data in a meaningful way and has made it possible to apply homological ideas to a wide variety of empirical data sets \cite{otter2017, Kaczynski2004}. PH is helpful for characterizing the ``shape'' of data, and the myriad applications of it include protein structure \cite{Gameiro2015, doi:10.1002/cnm.2655, kovacev2016, doi:10.1093/bib/bbs077}, DNA structure \cite{Emmett:2016:MTC:2954721.2954838}, neuronal structure \cite{Kanari2016QuantifyingTI}, computer vision \cite{Carlsson2008}, diurnal cycles in hurricanes \cite{tym2019}, inferring symbolic dynamics in chaotic systems \cite{ks2019}, spatial percolation problems \cite{speidel2018}, and many others. Additionally, combining machine-learning approaches with PH has also been very useful for several classification problems \cite{Adams:2017:PIS:3122009.3122017, Khasawneh2018, Yesilli2019, Cai2020}.

Because it is so natural to apply PH to the study of the shape of data, many successful applications of it have been to spatial networks. One particular area of interest has been the study of granular materials, because PH is able to effectively capture geometric properties of granular substances \cite{lia2018, Kramar2013, Buchet2018}. In addition to analyzing geometric information, PH methods are also able to describe multiscale spatial relationships. Many biological applications to proteins and DNA rely on the ability of PH to illuminate features at multiple scales, as multiscale structures and compositions of these molecules are extremely important to their functionality. PH has also been applied to larger-scale biological systems, including leaf-venation patterns \cite{leaf_optimization}, aggregation models \cite{topaz2015}, human migration \cite{Ignacio2019}, networks of blood vessels \cite{Byrne2019}, and the effects of psychoactive substances on brain activity \cite{Petri2014}. The recent review article \cite{batt2020} includes an extensive discussion of applications of PH to networks.

One confounding factor in the use of PH to study spatial networks is that although PH is able to capture information across scales, traditional distance-based PH constructions can have difficulty with applications in which differences in scale may not be meaningful. For example, in most applications to human geographical data, the difference in population density between urban and rural areas can dominate analyses that employ traditional PH constructions, and they thereby miss signals that do not rely on this variation in density. In a recent paper \cite{feng2019}, we examined the shape of voting patterns in the state of California and found that traditional methods for computing PH are more likely to capture disparities in population than to detect the presence of interesting voting patterns. To address this issue, we developed two novel PH constructions --- one based on network adjacency and one based on the physical geometry of a map --- that were more successful at capturing these voting patterns. For a recent analysis of the difficulty of interpreting signal and noise in PH results, see \cite{Bubenik2020}. For approaches other than PH for analyzing maps while accounting for density variation, see \cite{Tobler1963, Gastner2004}.

In the present paper, we apply our new PH constructions to a variety of spatial complex systems to demonstrate their usefulness across many domains. We show that these methods are well-adapted to capturing interesting structural properties of spatial networks and can thereby yield new insights into such networks, especially with respect to their global structure. Our examples include several synthetic graph models and dynamics on them, city street networks (which we compare both within a city and across different cities), snowflakes, and webs spun by spiders under the influence of various psychotropic substances.

Our paper proceeds as follows. In Section~\ref{sec:methods}, we give technical background on PH and on our particular constructions. In Section~\ref{sec:app}, we discuss computational results from computing the PH of (1) several well-known models of synthetic networks and (2) a variety of empirical data sets from diverse applications. We conclude in Section~\ref{conc}. A public repository of the code that we use for our computations is available at \url{https://bitbucket.org/mhfeng/spatialtda/src/master/}.


\section{Methods}
\label{sec:methods}


\subsection{Computing Persistent Homology}
\label{ss:ph}

We now give a brief introduction to PH and tools for computing it. See \cite{otter2017, Zomorodian2005, ghrist2008} for more details. We begin by defining $k$-simplices and simplicial complexes. A $k$-simplex is $k$-dimensional polytope that is a convex hull of $k+1$ nodes. A face of a $k$-simplex is any subset (of dimension smaller than $k$) of the $k$-simplex that is itself a simplex. A simplicial complex $K$ is a set of simplices that satisfy the following requirements: (1) if $\sigma \in K$ is a $k$-simplex, then every face of $\sigma$ is in $K$; and (2) if $\sigma$ and $\tau$ are simplices in $K$, then $\sigma \cap \tau$ is a face of both $\sigma$ and $\tau$.

Given a data set $X$, we construct a sequence $X_1 \subseteq X_2 \subseteq \cdots \subseteq X_l$ of simplicial complexes of some fixed maximum dimension.
We call the sequence $\{X_i\}$ a ``filtered simplicial complex'', and we call each $X_i$ a ``subcomplex'' of the filtered simplicial complex. We equip each relation $X_i \subseteq X_{i+1}$ with an inclusion map. The filtered simplicial complex, along with its inclusion maps and the chain and boundary maps of each subcomplex, constitutes a ``persistence complex''. The inclusion maps $X_i \hookrightarrow X_j$ induce a map $f_{i,j}\!: H_m(X_i) \to H_m(X_j)$ between homology groups. The map $f_{i,j}$ allows us to track an element of $H_m(X_i)$ (the $m$th homology group of the subcomplex $X_i$) to an element of $H_m(X_j)$. The $m$th homologies of the persistence complex are given by the pair 
\begin{equation}
	\left(\left\{ H_m(X_i) \right\}_{1 \leq i \leq l}, \left\{f_{i,j}\right\}_{1\leq i \leq j \leq l} \right)\,,
\end{equation}
and we call them the ``$m$th persistent homology'' of $X$. We refer to the collection of all $m$th persistent homologies as the ``persistent homology'' (PH) of $X$.

Consider a generator $x \in H_m(X_i)$ for some $m$ and $i$. If $x$ is not in the image of $f_{i-1,i}$, we say that $x$ is ``born'' at time $i$. Correspondingly, if $x \in H_m(X_i)$ and $f_{i,i+1}(x) = 0 \in H_m(X_{i+1})$, we say that $x$ ``dies'' at time $i+1$. If for every $j \leq l$, we have that $f_{i,j}(x) \neq 0$, then we say that $x$ never dies, and we assign a death time of $\infty$ to the element $x$. For each element $x$ of the PH of $X$, there is a birth time $b_x$ and a death time $d_x$, and the collection of intervals $\{[b_x, d_x)\}$ is the ``barcode'' of $X$. Generators with longer associated half-open intervals $[b_x, d_x)$ are more persistent. It is traditional to construe more-persistent intervals as better indicators of a signal and construe less-persistent intervals as noise, although recent work (including our own \cite{feng2019, stolz2017}) indicates that it is not always possible to interpret persistence in this way.

The collection of features in each $H_m(X_i)$ describes the topological properties of the filtration $\{X_i\}$. Intuitively, each feature in $H_m$ corresponds to some $m$-dimensional void. In $H_0$, features are connected components; in $H_1$, features are loops. By considering the PHs of $\{X_i\}$, we can examine how the connectedness of $\{X_i\}$ changes for each step of the filtration in each dimension. For example, a short-persistence feature in $H_0$ is a connected component that appears and combines quickly with another component. PH records all features and their persistences, allowing us to take a global view of topological changes in each filtration step of $\{X_i\}$.

In the present paper, we use the software package {\sc Gudhi} \cite{gudhi:urm,gudhi:FilteredComplexes} to compute PH of the filtered simplicial complex $\{X_i\}$. We construct $\{X_i\}$ from $X$ using two different constructions, which we developed recently in a paper on voting data \cite{feng2019}.


\subsection{Adjacency Construction of PH}
\label{ss:adj}

\begin{figure}[t]
	\centering
	\includegraphics[width=.48\textwidth]{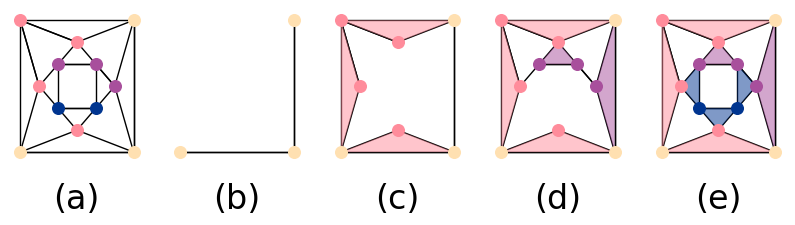}
\caption{\label{fig:adj_construction}{We illustrate an adjacency construction of persistent homology (PH) on (a) a planar graph, whose nodes we color according to a function value from yellow to dark blue. At each filtration step (see panels (b)--(e)), we add all nodes with a given range of function values. We also add any edges between these new nodes, as well as any edges between these new nodes and existing nodes, and we fill in any triangles that form. Only cycles of length three form triangles, so the graph in panel (a) yields 5 infinite-length features in $H_1$ [as one can see from the five holes that remain in panel (e)].}
}
\end{figure}

We now describe a way to construct a filtered simplicial complex based on network adjacencies. We consider a network in the form of a graph $(V,E)$, with numerical data $f(v)$ associated with each node $v$. For a given filtration step $X_i$, let the $0$-simplices of $X_i$ be given by $v \in V$ such that $f(v) \leq \epsilon$ for some value $\epsilon$. For any edge $(u,v) \in E$, if $u \in X_i$ and $v\in X_i$, we add $(u,v)$ to $X_i$. Finally, to $X_i$, we add all triangles $(u,v,w)$ such that $(u,v)$, $(v,w)$, and $(u,w)$ are in $X_i$. We repeat this process for $X_{i+1}$, but now we use a larger value of $\epsilon$. By construction, each $X_i \subseteq X_{i+1}$, and we have a valid filtered simplicial complex. See Fig.~\ref{fig:adj_construction} for an illustration of such a filtered simplicial complex. 

This adjacency construction tracks topological changes in a network as it grows. The homology group $H_m(X_0)$ characterizes the topology of the first filtration step. As one adds more nodes, edges, and faces to the simplicial complex, the topology changes and is recorded in $H_m(X_i)$. By choosing $f$ carefully, we can control which subset of a network exists in the first filtration step, and we can also control how the network expands. For example, in \cite{feng2019}, by attaching voting data to a network of precincts of a county, we used the adjacency construction to examine how the topology of a county changes as one considers precincts with an increasingly wide range of voting preferences.

In some of our applications, we use an alternate adjacency construction in which we associate data $g(u,v)$ to each edge $(u,v)$, instead of to the nodes. This construction differs from the one above only in that we define the function $\tilde{f}(v) = \min_{\{u: (u,v) \in E\}} g(u,v)$. We then proceed with the above adjacency construction, but we substitute $\tilde{f}$ for $f$. We recently introduced our main adjacency construction in \cite{feng2019}, and we introduce this adaptation of it to edge-based data in the present work.

\subsection{Level-Set Construction of PH}
\label{ss:ls}

\begin{figure}[htbp!]
	\centering
	\begin{subfigure}{.15\textwidth}
		\includegraphics[width=\textwidth]{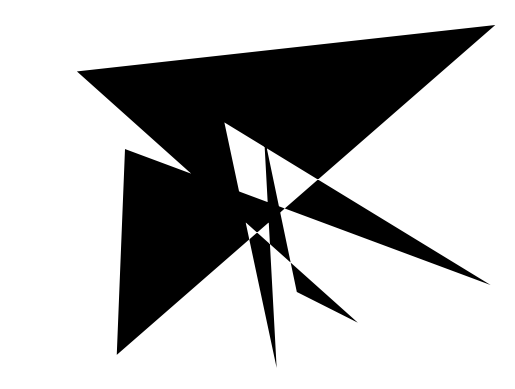}
		\caption{}
	\end{subfigure}
	\begin{subfigure}{.15\textwidth}
		\includegraphics[width=\textwidth]{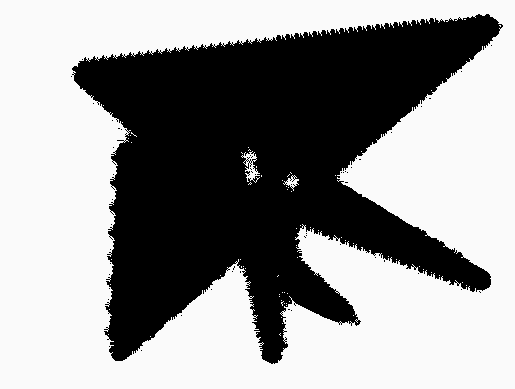}
		\caption{}
	\end{subfigure}
	\begin{subfigure}{.15\textwidth}
		\includegraphics[width=\textwidth]{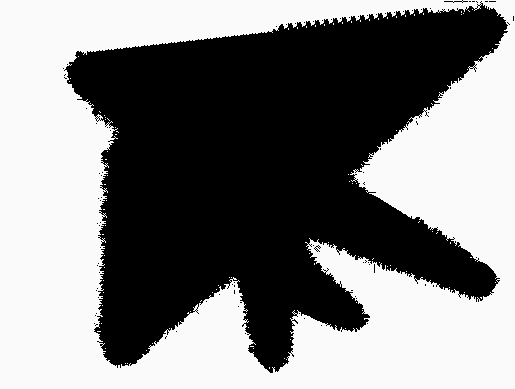}
		\caption{}
	\end{subfigure}
\caption{\label{fig:ls_example} Illustration of level-set dynamics. Starting from (a) an initial black-and-white image, we apply level-set evolution \eqref{level} for several steps to obtain the image in (b) and then the one in (c). In these images, the white space in the center of the image shrinks until eventually it is completely covered by the expanding black surface.
}
\end{figure}

The other PH construction that we use (again see \cite{feng2019} for details) involves describing data as a manifold, rather than as a graph. Let $M$ denote a two-dimensional (2D) manifold, such as data in an image format. We consider the boundary $\Gamma$ of $M$ and construct a sequence
\begin{equation*}
	M_0 \subseteq M_1 \subseteq \cdots \subseteq M_n
\end{equation*}	
of manifolds, where at each time step, we evolve the boundary $\Gamma_t$ of $M_t$ outward according to the level-set equation. (See \cite{osher2003} for a thorough exposition of the level-set equation and level-set dynamics.) That is, for a manifold $M$ that is embedded in $\mathbb{R}^2$, we define a function $\phi(\vec{x},t)\!: \mathbb{R}^2 \times \mathbb{R} \to \mathbb{R}$, where $\phi(\vec{x},t)$ is the signed distance function from $\vec{x}$ to $\Gamma_t$ at time $t$. We propagate $\Gamma_t$ outward at velocity $v$ using the equation
\begin{equation} \label{level}
	\frac{\partial \phi}{\partial t} = v |\nabla \phi |
\end{equation}
until all homological features have died. Because this evolution gives a signed distance function at each time step $t$, we take $M_{t}$ to be the set of points $\vec{x}$ such that $\phi(\vec{x},t) > 0$. (This corresponds to points inside the boundary $\Gamma_t$.) We show an example of this evolution in Fig.~\ref{fig:ls_example}. Throughout this paper, we use $v=1$. Different values of $v$ cause the level set to evolve faster (if $v > 1$) or slower (if $v < 1$), resulting in a different number of time steps (and hence a different number of filtration steps) in our evolution. However, we obtain the same homological features, although with different birth times and death times. If $v$ is sufficiently large, it is possible that all features would have the same birth and death time, such that no features would occur after the first filtration step. When $v=0$, there is no evolution.

\begin{figure}[htbp!]
	\centering
	\begin{subfigure}{.22\textwidth}
		\includegraphics[width=\textwidth]{figures/ls/test-2}
		\caption{}
	\end{subfigure}
	\begin{subfigure}{.22\textwidth}
		\includegraphics[width=\textwidth]{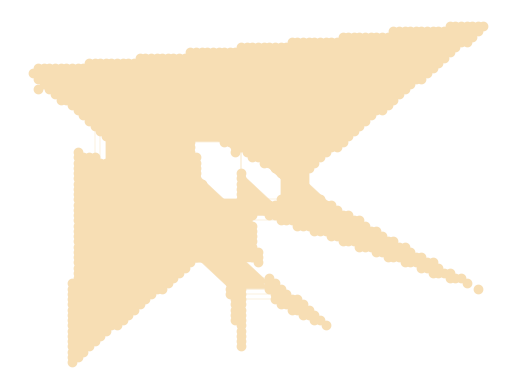}
		\caption{}
	\end{subfigure}
	\begin{subfigure}{.22\textwidth}
		\includegraphics[width=\textwidth]{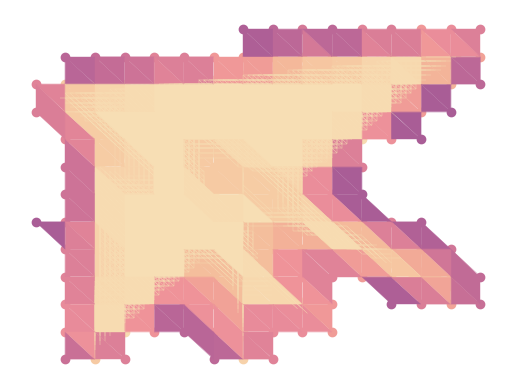}
		\caption{}
	\end{subfigure}
	\begin{subfigure}{.22\textwidth}
		\includegraphics[width=\textwidth]{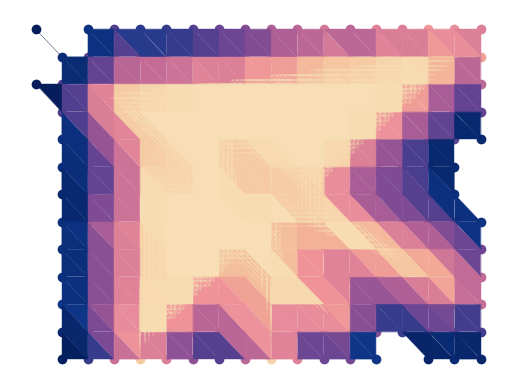}
		\caption{}
	\end{subfigure}
\caption{\label{fig:ls_tri} Illustration of a level-set adjacency construction of PH. In (a), we show a synthetic image that we use as an initial manifold for level-set evolution. In (b)--(d), we show various filtration steps of the filtered simplicial complex that we generate by performing a level-set evolution on the image in panel (a). Panel (b) shows the simplicial complex that we obtain by overlaying the image in panel (a) on a triangular grid. In panels (c) and (d), we add new nodes, edges, and triangles to the image as it evolves outward. Darker colors indicate simplices that enter the filtration at a later time step.
}
\end{figure}

By imposing $\{M_i\}$ over a triangular grid of points, as described in \cite{feng2019}, we obtain a corresponding simplicial complex $X_i$ for each $M_i$. In Fig.~\ref{fig:ls_tri}, we give a visualization of this simplicial complex. We construct this level-set complex using a polygon whose points we choose uniformly at random from $[0,1] \times [0,1]$ as an initial synthetic image. Because the level-set equation \eqref{level} evolves continually outward, we automatically satisfy that condition that $X_i \subseteq X_{i+1}$, so $\{X_i\}$ is a filtered simplicial complex. Our implementation of the level-set method works with any black-and-white image (or any image that one formulate as a piecewise-constant function $\mathbb{R}^2 \to \{0,1\}$). We expect our level-set approach to capture information about $H_0$ and $H_1$ for any such image. The level-set approach also captures geometric information, which can be useful for some applications; however, this may make it difficult to capture information about holes that are visually irregular. Throughout this paper, we compare images that have roughly the same resolutions, where we take the image resolution from raw image data. Because image size should primarily affect the computation time of our level-set approach --- but not the order in which features appear and disappear as an image evolves --- we expect that it is possible to adapt our level-set construction when comparing images of different resolutions. Possible approaches for such an adaptation include normalizing image sizes or adjusting the resolution of the triangular grid that one uses for each image.

\section{Applications}
\label{sec:app}

We now discuss applications of PH to both synthetic networks and empirical spatial networks from a diverse variety of applications.


\subsection{Synthetic Networks}
\label{ss:random}

In this subsection, we discuss applications of our adjacency PH construction to a dynamical process on synthetic networks in which space plays an important role. 
For each network $(V, E)$, we run the Watts threshold model (WTM) \cite{watts2002} on it. Given a graph, we select a fraction $\rho_0 = 0.05$ of its nodes uniformly at random to be ``infected'' at time $0$. At each time step, we then we compute the fraction of each node's neighbors that are infected. (That is, we synchronously update the states of the nodes \cite{porter2016}.) If the fraction of a node's neighbors that are infected meets or exceeds a threshold (in our case, the threshold is $\varpi = 0.18$ for all nodes), the node becomes infected. We take this implementation of the WTM to be the generator of a function $f\!: V \to \mathbb{N}$ \footnote{We use the convention that $\mathbb{N}$ includes $0$.}, where $f(v)$ is the time at which node $v$ becomes infected. We say that infected nodes are in the set $I$. If $v$ never becomes infected, we set $f(V) = \max_{v \in I} f(v) + 1$, so that we eventually add all nodes to a filtered simplicial complex. The resulting filtered simplicial complex consists of the subgraphs that are generated by $I$ at each time step. See \cite{taylor2015,barb2018,yingthesis} for studies of the WTM on spatial networks. 

We use parameter values of $\rho_0 = 0.05$ and $\varpi = 0.18$ throughout this section. We expect changes in $\rho$ and $\varpi$ to affect the birth times and death times of features in a filtered simplicial complex. Using a different value of $\rho_0$ entails considering a different fraction of initially infected notes. Therefore, a larger value of $\rho_0$ yields a larger simplicial complex at the first filtration step, and smaller value of $\rho_0$ yields a smaller simplicial complex. Using a larger value of $\varpi$ results in fewer nodes becoming infected at each time step, and it thus takes more filtration steps before the simplicial complex stops growing. Because our underlying graph is the same for any choice of values of $\rho$ and $\varpi$, we do not expect changes in the homology of the last filtration step, unless $\rho$ or $\varpi$ are sufficiently small such that some nodes in a graph never become infected. However, one can certainly obtain a different PH for different values of $\rho$ or $\varpi$, as nodes and edges can join the filtered simplicial complex at different times and (more importantly) in different orders.

\begin{figure}[htbp!]
	\centering
	\includegraphics[width=.45\textwidth]{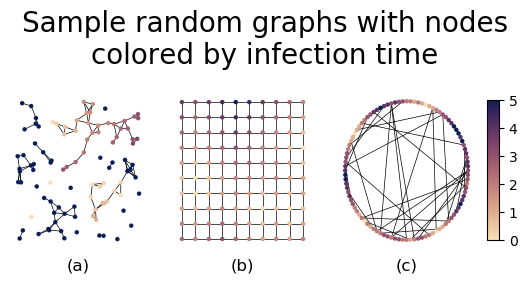}
	\caption{An instance of each of our synthetic networks with Watts threshold model (WTM) dynamics on it. The corresponding persistence diagrams (PDs) are in Figs.~\ref{fig:rgg_pd},~\ref{fig:lattice_pd}, and~\ref{fig:ws_pd}. We color the nodes based on the time that they become infected. The three types of 
	synthetic networks are (a) a random geometric graph, (b) a square lattice network, and (c) a Watts--Strogatz small-world network.
	\label{fig:random_graphs}
	}
\end{figure}

We examine topological changes in the infected subgraph of three different types of synthetic networks (see Fig.~\ref{fig:random_graphs}). We first examine random geometric graphics (RGGs) \cite{penrose-rgg}. For each instance of an RGG, we pick $100$ nodes uniformly at random from the unit square. If the Euclidean distance between two nodes is less than or equal to $0.1$, we add an edge between them [see Fig.~\ref{fig:random_graphs}(a)]. Our second type of synthetic network is a square lattice with $100$ nodes. We arrange the $100$ nodes in a $10 \times 10$ grid on the unit square, and we then connect the nodes along the grid lines [see Fig.~\ref{fig:random_graphs}(b)]. Our third type of synthetic network is a Watts--Strogatz (WS) small-world network \cite{watts_strogatz_1998,scholarsmall}. We begin with a ring of $100$ nodes, and we then connect each node to its $k=2$ nearest neighbors on each side. 
We then rewire each edge uniformly at random with a probability of $p=0.1$ using the implementation of the WS model in {\sc NetworkX} \cite{SciPyProceedings_11}. In this version of the WS model, one removes each rewired edge before replacing it with a new edge. We show an example of a WS graph in Fig..~\ref{fig:random_graphs}(c).

For each type of synthetic network, we consider $100$ instances, which we generate using {\sc NetworkX}. For the RGG and WS networks, each instance is a different graph; the square lattice network is deterministic. For all three types of networks, each instance has a different initial set of infected nodes. We show visualizations of each of these types of networks (with WTM dynamics on it) in Fig.~\ref{fig:random_graphs}.

Our adjacency construction begins by selecting the initially infected nodes and the edges between them of a network to create an infected subgraph that we call an ``infection network''. As an infection spreads, we add more nodes and edges to the infection network until eventually we have added all nodes and edges to it.

\begin{figure}[htbp!]
	\centering
	\includegraphics[width=.35\textwidth]{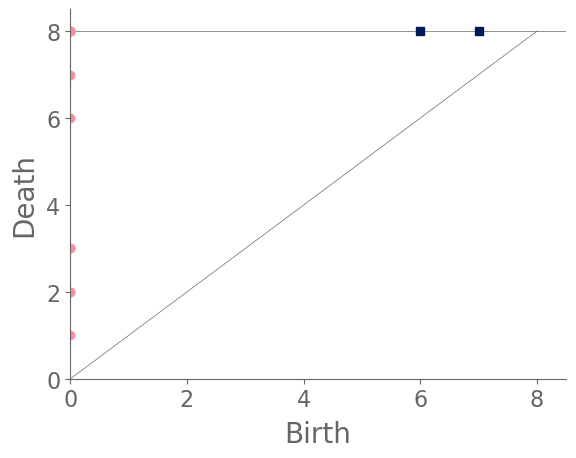}
	\caption{\label{fig:rgg_pd} The PD for an instance of the WTM on an RGG. We plot each feature as a point on the PD, for which the horizontal coordinate represents the birth time and the vertical coordinate represents the death time. We plot features with infinite persistence (i.e., features that do not die within the range of filtration parameters that use for a PH computation) on a horizontal line at the top of the PD. We plot features in $H_0$ (which indicates the connected components) as pink circles, and we plot features in $H_1$ (which indicates the one-dimensional holes) as dark-blue squares.
	}
\end{figure}

Examining the PHs of the RGGs (see Fig.~\ref{fig:rgg_pd}), we see for our parameter values that an infection network tends to have several connected components, resulting in a large number of features in $H_0$. However, because of the spreading behavior of the WTM, new nodes can become infected only via their infected neighbors. Because features in $H_0$ record connected components of a graph, new infected nodes join existing connected components. Therefore, features can only be born at time $0$ or in the last step, which is when we add all remaining uninfected nodes to our filtered simplicial complex. 
By contrast, features in $H_1$ are relatively rare, as most cycles that occur in an RGG are filled because of the uniform probability distribution of the node locations.

\begin{figure}[htbp!]
	\centering
	\includegraphics[width=.35\textwidth]{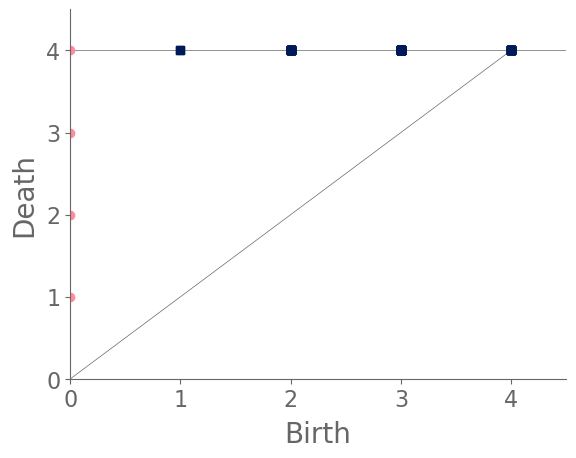}
	\caption{\label{fig:lattice_pd} The PD for an instance of the WTM on a square lattice network.}
\end{figure}

For a square lattice network (see Fig.~\ref{fig:lattice_pd} for a PD of the WTM on such a network), we first note that there is only a single infinite-length feature in $H_0$, as the final infection network necessarily consists of a single connected component. Consequently, $H_0$ consists of a set of features that are born at time $0$ and eventually merge (and therefore die), resulting in a single infinite-length feature. Additionally, there are a constant number (81, to be precise) of features in $H_1$, because when we construct a simplicial complex, every grid cell of the lattice is a feature in $H_1$ at the last filtration step. However, these features can be born at a variety of times, as the filtration does not include every lattice cell until every node of the graph has entered the filtration.

\begin{figure}[htbp!]
	\centering
	\includegraphics[width=.35\textwidth]{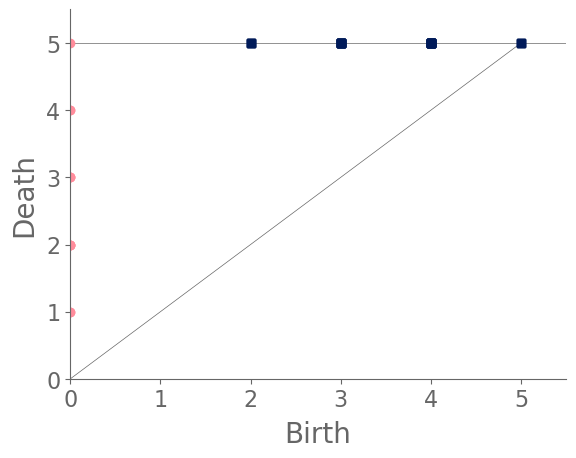}
	\caption{\label{fig:ws_pd} The PD for an instance of the WTM on a WS network.}
\end{figure}

From Fig.~\ref{fig:ws_pd}, we see that a WS small-world network also eventually has an infection network that consists of a single connected component. However, the WS networks consistently have more features in $H_1$ than the RGG networks, because the former's (non-geometric) shortcut edges usually result in splitting an existing cycle (and hence a feature in $H_1$) into two cycles.

We summarize our observations about the various synthetic networks in Table~\ref{tab:random_comp}, in which we give the means and standard deviations of the number of features during the temporal evolution of the WTM in each type of synthetic graph. Our counts include features that appear at any time during the WTM dynamics.

\begin{table}[htbp!]
	\caption{\label{tab:random_comp}Means and standard deviations of the numbers of features in $H_0$ and $H_1$ during the temporal evolution of the WTM across all instantiations of each type of synthetic graph. We conduct 100 simulations for each network model, and our counts include features that appear at any time during the WTM dynamics.
	}
	\begin{tabular}{@{}lcccc@{}}\toprule
	& Mean ($H_0$) & STD ($H_0$) & Mean ($H_1$) & STD ($H_1$) \\ \midrule
	RGG & 23.16 & 3.1897 & 1.2 & 1.0 \\
	Square lattice & 4.56 & 0.5886 & 81 & 0 \\
	WS & 8.29 & 2.0214 & 26.95 & 5.2314 \\ \bottomrule
	\end{tabular}
\end{table}


\subsection{Street Networks in Cities} 
\label{ss:cities}

The field of urban analytics has grown rapidly in the last several years \cite{barthelemy2018,barthelemy2019,pumain2020}, Increasingly powerful computational tools have allowed researchers to characterize cities in terms of their street networks \cite{boeing2019b}, and a variety of approaches from network analysis have been applied to the study of urban street networks \cite{boeing2018multi, Barthelemy2017, cardillo2006, Batty, Ahmed2014, Wu2018, thompson2020}. In the present subsection, we use city street networks as base manifolds in our level-set construction, and we thereby characterize cities based on their PHs. We use these PHs to compare city morphologies both within a single city and across a variety of cities.

We use our level-set construction to obtain topological descriptors in the form of persistence for city street networks. We then use these city persistences to compare (1) different regions of the same city and (2) different cities to each other. We obtain all of our city street networks with the software package {\sc OSMnx} \cite{boeing_2017} using latitude--longitude coordinates and taking a $1$ km block that is centered at specified coordinates. In each example, we indicate how we choose these coordinates.

The first filtration step of a filtered simplicial complex that results from our PH construction consists only of the streets in a network. As we increase the filtration time, we slowly add city blocks to the complex, and the topology changes as those blocks are filled in. More regular city blocks are more likely to be filled in without creating any new homological features, and larger blocks take longer to be filled in. Our construction is thereby able to capture information about the size and regularity of city blocks. The existence of dead ends tends to lead to the ``pinching'' of blocks into multiple homological features --- as dead ends expand, they lengthen and eventually meet with nearby streets, cutting through blocks in the process --- so our approach also yields information about dead ends.


\subsubsection{Comparing Different Regions of the Same City}
\label{sss:within_cities}

\begin{figure}[htbp]
	\centering
	\begin{subfigure}[t]{0.2\textwidth}
		\centering
		\hspace{.11\textwidth}\includegraphics[width=.88\textwidth, height=.78\textwidth]{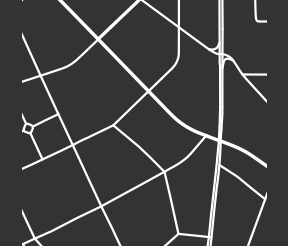}\\
		\includegraphics[width=\textwidth, height=.82\textwidth]{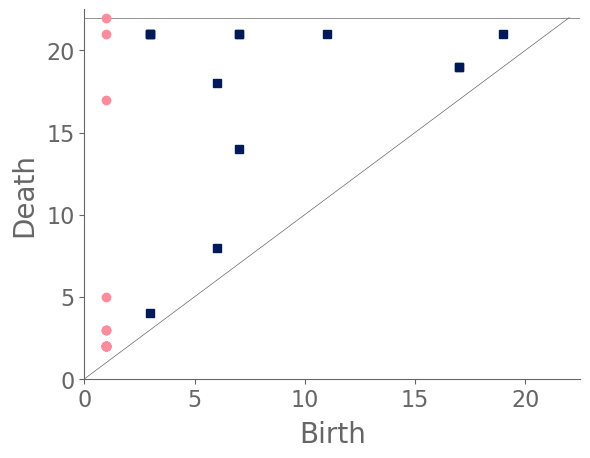}
		\caption{}
	\end{subfigure}
	\begin{subfigure}[t]{0.2\textwidth}
		\centering
		\hspace{.11\textwidth}\includegraphics[width=.88\textwidth, height=.78\textwidth]{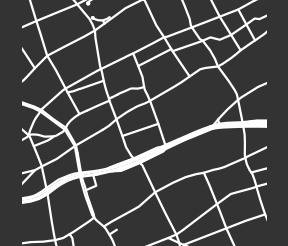}\\
		\includegraphics[width=\textwidth, height=.82\textwidth]{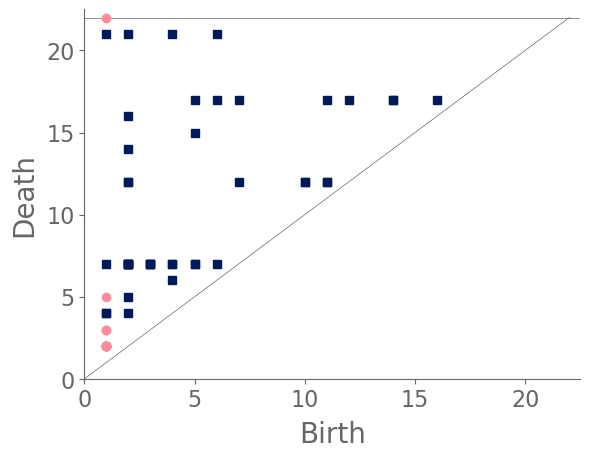}
		\caption{}
	\end{subfigure}
	\caption{\label{fig:shanghai_sn}Two sampled street networks from (a) Pudong New Area and (b) Zhabei district. [We generated both maps using {\sc OSMnx} \cite{boeing_2017}.]
	}
\end{figure}

We sample 169 points from the city of Shanghai using a {\sc shapefile} of Shanghai's administrative-district boundaries that we downloaded from {\sc ArcGIS} \cite{shanghai_map}. From the {\sc shapefile}, we obtain a bounding box for each point. We sample uniformly within this bounding box, discarding points that do not lie within the polygonal district geometry that is defined in the {\sc shapefile}. We stop sampling when we reach the desired number of points.
In total, we sample ten points from each administrative district, and we also include nine historical landmarks with coordinates from Google Maps \cite{googlemaps}. In Fig.~\ref{fig:shanghai_sn}, we show maps and their associated PDs for two examples.

After computing PH (in the form of a PD) for each map, we compute the bottleneck distance between each pair of maps. Bottleneck distance is a metric that is defined on the space of PDs. It gives the shortest distance $d$ for which there exists a perfect matching between the points of the two PDs (along with all diagonal points), such that any pair of matched points are at most a distance $d$ from each other, where we use the supremum norm in $\mathbb{R}^2$ to compute the distance between points. 
Once we have pairwise bottleneck distances between PDs, we perform average-linkage hierarchical clustering into three clusters. (We chose to have three clusters based on looking at the dendrogram.) We can replace our metric with a different metric (such as a Wasserstein distance \cite{Kerber2017}) on PDs or cluster our PDs using a different clustering algorithm. We do not discuss the impact that such choices may have on our results, although we note in passing that we performed $k$-medoids clustering \cite{Park2009} for our case study of Shanghai and obtained qualitatively similar results.

\begin{figure}[htbp]
	\centering
	\includegraphics[width=.45\textwidth]{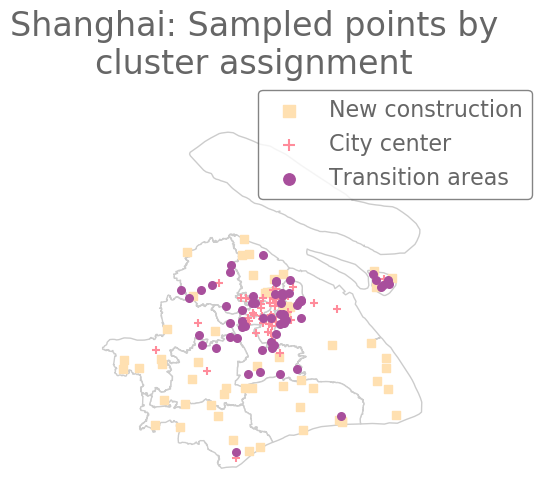}
	\caption{\label{fig:shanghai_clusters} Sampled points in Shanghai. We color these points according to their cluster assignment from average-linking hierarchical clustering of areas of Shanghai into three clusters.
	}
\end{figure}

\begin{figure}[htbp]
	\centering
	\includegraphics[width=.45\textwidth]{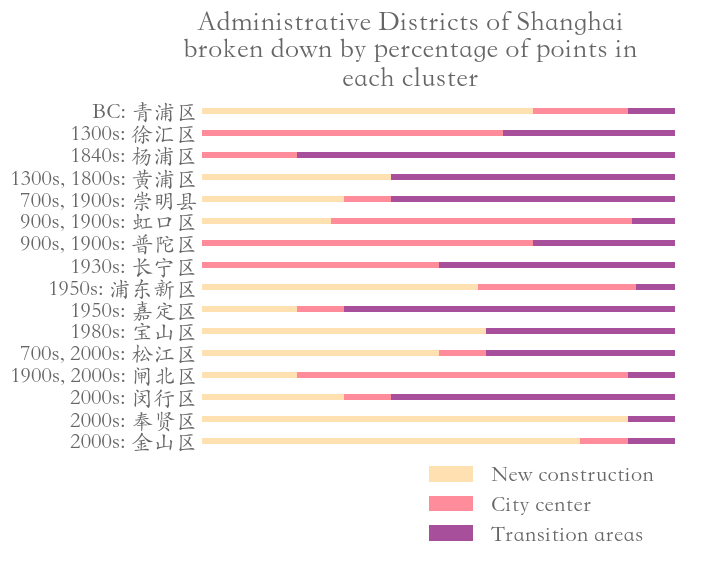}
	\caption{\label{fig:shanghai_nbd_pie}
	Breakdown of administrative districts in Shanghai into our three clusters. (We order the districts roughly by their year of development.) Most of the older districts have a larger percentage of points that are assigned to the ``City center" cluster, whereas the points in the ``Transition areas'' cluster tend to occur in districts that included development in the 19th and early 20th centuries. The ``New construction" cluster is the most common assignment for administrative districts from the 1950s or later.
	}
\end{figure}

In Fig.~\ref{fig:shanghai_clusters}, we show the sampled points (which we color according to their cluster assignment). We observe that the three clusters consist largely of historical areas (``City center''), concession-era areas (``Transition areas''), and modern areas (``New construction''). In Fig.~\ref{fig:shanghai_nbd_pie}, we show administrative districts along with the year that they were constructed. We break them down by the percentage of the sample points that are in each cluster.

\subsubsection{Comparing Street Networks from Different Cities} 
\label{sss:across_cities}

We continue our analysis of cities by characterizing and comparing the structures of street networks of 306 cities across the globe. We downloaded latitude and longitude coordinates from SimpleMaps \cite{simplemaps} and selected all cities with a population of at least $1.5$ million people. Given these latitude and longitude coordinates, we use {\sc OSMnx} \cite{boeing_2017} to obtain street networks. We then compute PH for each city and cluster their PDs using average-linkage hierarchical clustering with three clusters. We sometimes refer to a city in a given cluster as a city of a certain ``type''. Our results depend on the specific latitude and longitude coordinates in our downloaded data set. Accordingly, our results are influenced by the particular location of a city's coordinates, which are the standard ones in SimpleMaps.

In the following paragraphs, we describe our clusters of cities. We define ``blocks'' to be the cells of a planar street graph. Although our level-set construction for computing PH is not designed explicitly to characterize blocks, we take advantage of the fact that our level-set construction takes the set of streets as its initial manifold. As the streets expand outward according to the level-set equation \eqref{level}, they fill in the blocks. Larger blocks take longer to fill in, and blocks fill in more evenly when they are closer to circular in shape. Roughly, we characterize block sizes based on the death times of features in $H_1$: small sizes correspond to early death times (specifically, less than 10), medium sizes correspond to death times between 10 and 15, and large sizes correspond to late death times (specifically, more than 15). We also designate blocks as ``regular'' (when they are close to a regular convex polygon) or ``irregular'' (for blocks that do not resemble a rectangle or some other regular convex polygon). If a block is very irregular, then as its streets expand, it is possible that narrow parts of the block will shrink and close off, such that the block segments into smaller blocks. We refer to this phenomenon as ``pinching''. Our three main clusters are dominated by (1) gridlike cities; (2) cities with gridlike patches that are interspersed with larger, non-gridlike blocks; and (3) cities that have a large number of non-gridlike structures (specifically, dead ends or large holes) that interrupt other structures. We use the term ``interrupted grid'' for cities that are either (1) mostly gridlike with some patches that are not gridlike or (2) consist of patches of disparate grids that are stitched together (with other features between them).

Our first major cluster has 99 cities and is dominated by cities with small, gridlike blocks. All regions of the world have some cities of this type, but North America has the largest percentage (relative to all of the cities that we sample from that continent) of these gridlike cities and Europe has the smallest percentage of them. The block sizes in this cluster tend to be small or of medium size, resulting in filtrations whose maximum filtration value tends to be small in comparison to cities in the other two clusters. In the PDs, we also observe that the distributions of death times of features in $H_1$ tends to be close to uniform and over a small range. 
Such distributions occur because these gridlike cities tend to have even distributions of block sizes, even though they include some areas with slightly smaller and/or slightly larger grid sizes. They do not have large blocks, so they do not have features in $H_1$ with late death times.

\begin{figure}[htbp]
	\centering
	\hspace{.03\textwidth}\includegraphics[width=.4\textwidth, height=.369\textwidth]{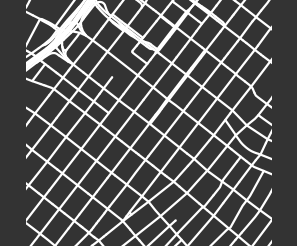}\\
	\includegraphics[width=.45\textwidth, height=.369\textwidth]{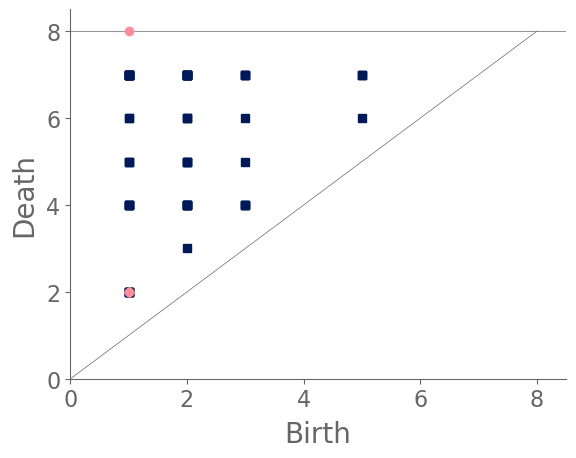}
	\caption{\label{fig:intra_cluster_one} Cities in our first major cluster have gridlike street layouts. One example of such a city is Los Angeles, which we show in this figure. We show its street network in the top row and its associated PD in the bottom row.
	}
\end{figure}

\begin{figure}[htbp]
	\centering
	\begin{subfigure}[t]{0.23\textwidth}
		\centering
		\hspace{.11\textwidth}\includegraphics[width=.88\textwidth, height=.82\textwidth]{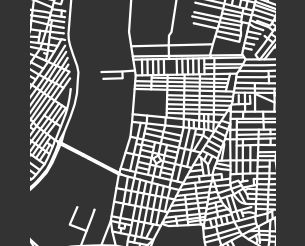}\\
		\includegraphics[width=\textwidth, height=.82\textwidth]{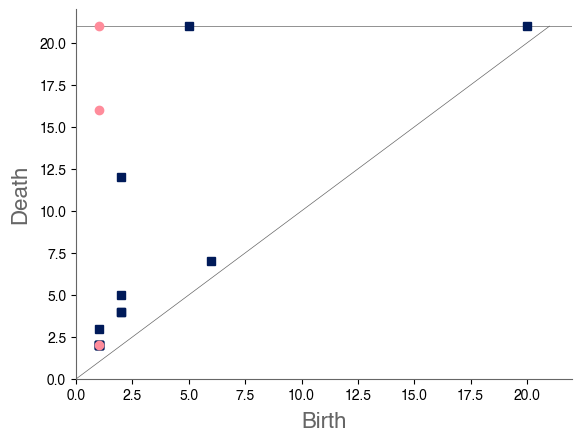}
		\caption{}
	\end{subfigure}
	\begin{subfigure}[t]{0.23\textwidth}
		\centering
		\hspace{.11\textwidth}\includegraphics[width=.88\textwidth, height=.82\textwidth]{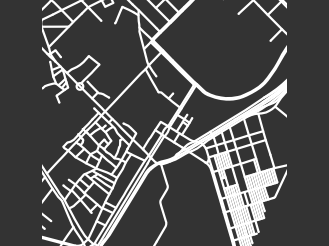}\\
		\includegraphics[width=\textwidth, height=.82\textwidth]{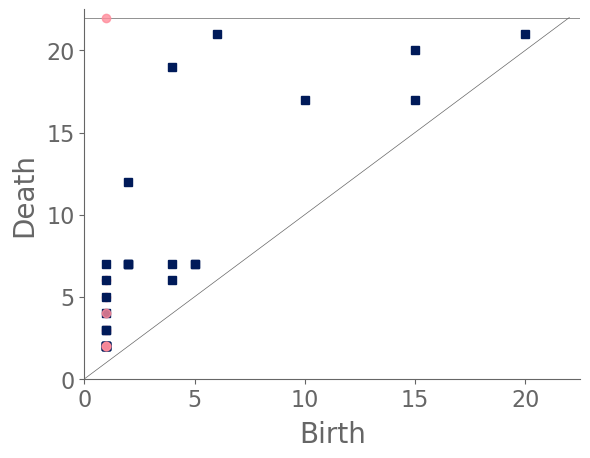}
		\caption{}
	\end{subfigure}
	\caption{\label{fig:intra_cluster_two} Cities in our second major cluster have patches of gridlike structure that are mixed with large blocks. As examples of cities in this cluster, we show (a) Aleppo and (b) Barcelona. We show their street networks in the top row and their associated PDs in the bottom row. Aleppo illustrates the idea of having holes in a large grid and is an example of a city in the first subcluster of cluster two. Barcelona, which is in the second subcluster, is an example of a city with small patches of gridlike structure.
	}
\end{figure}

Our second major cluster has cities with patches of grids that are interspersed with structures that are not gridlike. This cluster, with 149 cities, is the largest of our three clusters. The PDs in this cluster tend to have larger maximum death times than the PDs for the cities in our first cluster. In the PDs, gridlike blocks yield collections of features in $H_1$ with early death times; and the larger, non-gridlike structures yield features in $H_1$ with late death times. The non-gridlike areas in these cities tend to have fairly regular shapes, resulting in a relatively small number of features in $H_1$ with late birth times. Such late-birth-time features usually correspond to the pinching of blocks, which can occur either via dead ends or via shape irregularities. By examining the dendrogram from our hierarchical clustering, we can further separate the second cluster into two subclusters, which we show in Fig.~\ref{fig:intra_cluster_two}. The first of these subclusters consists mostly of cities that have large patches of gridlike structure, with a small number of large blocks that interrupt the grids. The PDs for cities in this subcluster tend to have a large number of features in $H_1$ with early death times, and they tend to have only a small number of isolated features in $H_1$ with late death times. The second subcluster of our second major cluster consists mostly of cities with small patches of grids that are mixed with large irregular blocks. The PDs for cities in this subcluster tend to have a larger number of features in $H_1$ with late death times than is the case for the cities in the other subcluster of cluster two.

\begin{figure}[htbp!]
	\centering
	\begin{subfigure}[t]{0.23\textwidth}
		\centering
		\hspace{.11\textwidth}\includegraphics[width=.88\textwidth, height=.82\textwidth]{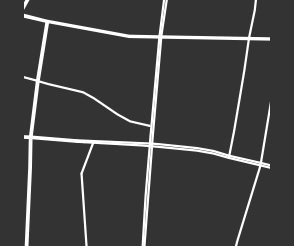}\\
		\includegraphics[width=\textwidth, height=.82\textwidth]{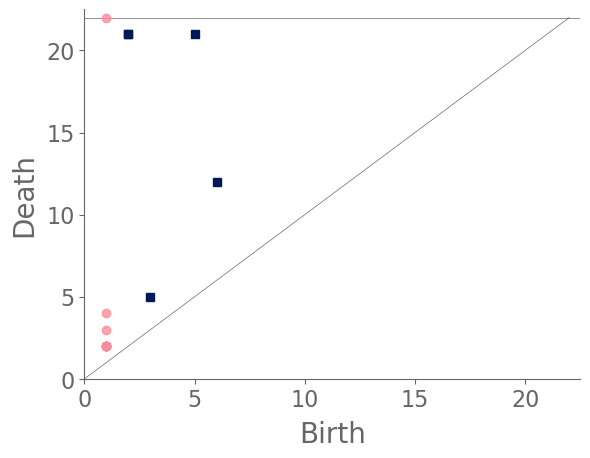}
		\caption{}
	\end{subfigure}
	\begin{subfigure}[t]{0.23\textwidth}
		\centering
		\hspace{.11\textwidth}\includegraphics[width=.88\textwidth, height=.82\textwidth]{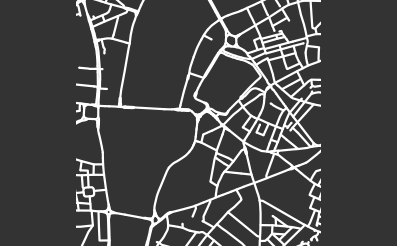}\\
		\includegraphics[width=\textwidth, height=.82\textwidth]{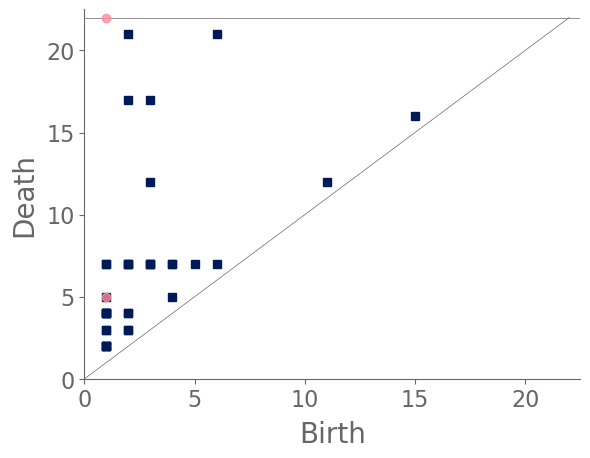}
		\caption{}
	\end{subfigure}
	\caption{\label{fig:intra_cluster_three} 
	Examples of cities in our third major cluster include (a) Nanyang and (b) London. We show their street networks in the top row and their associated PDs in the bottom row. Cities in our third major cluster include dead ends, irregular blocks, and obstructions. This leads to a large range of block sizes and hence to features in $H_1$ that have medium death times. Such features are rare in the other two major clusters. For example, Nanyang has several streets with obstructions, and London has dead ends and a broad distribution of block sizes.
	}
\end{figure}

Our third major cluster, with 58 cities, consists of cities with a large number of non-gridlike structures. In particular, many of these cities include a large number of dead ends, rectangular blocks that are not arranged in a grid, or both. We observe streets that do not continue through particular blocks (e.g., there is a street, it is obstructed, but then it continues after the obstruction), which leads to a mixture of block sizes even in areas of a city that tend to have regular blocks. We refer to these situations as ``obstructions''. The PDs of the cities in this cluster have a larger number of features in $H_1$ with medium death times (specifically, in the range 10--15), and many of these features are close to the diagonal. This is common when large blocks are pinched into several regions, as the smaller regions are born at the pinching time, rather than near the beginning of the filtration. Therefore, they do not survive long enough to have a late death time. By examining the dendrogram from our hierarchical clustering, we see two clear subclusters. However, one of these subclusters consists of only two cities (Beirut and Nanyang). The PDs of both of these cities are dominated by two features in $H_1$ with late death times, and they also have several features in $H_1$ with medium death times. In Fig.~\ref{fig:intra_cluster_three}, we show examples of cities in cluster three.

\begin{turnpage}
\begin{figure*}[htbp!] 
	\centering
	\includegraphics[width=1.4\textwidth]{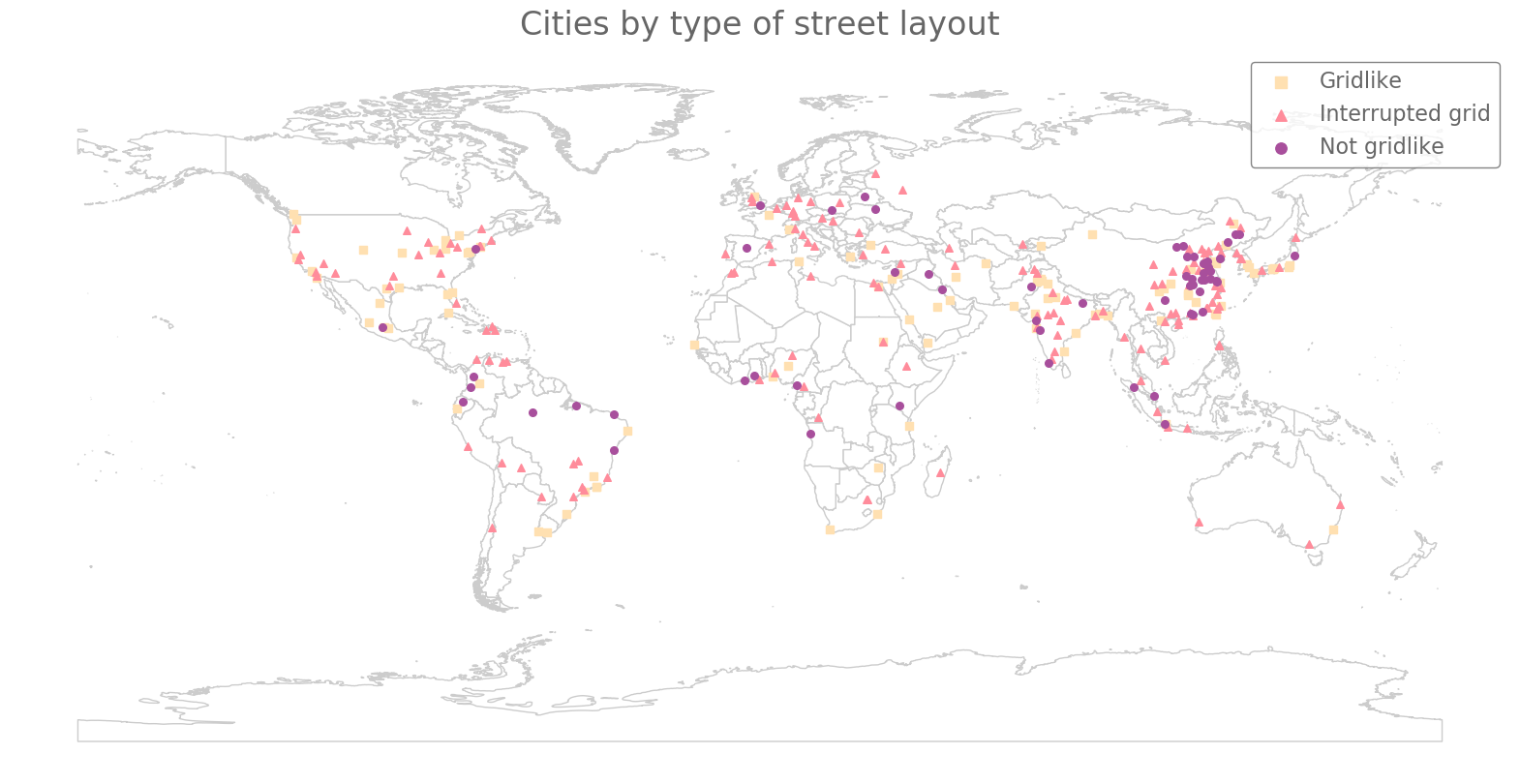}
	\caption{\label{fig:world_clusters} Cities colored by their cluster assignments from average-linkage hierarchical clustering of cities into three clusters.
	  [The {\sc shapefile} of the world map is from \cite{world_map}.]
	}
\end{figure*}
\end{turnpage}

\begin{figure}[htbp!] 
	\centering
	\includegraphics[width=.45\textwidth]{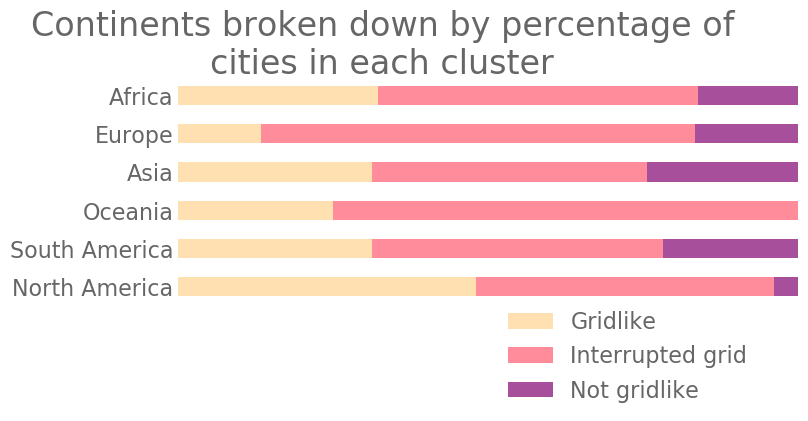}
	\caption{\label{fig:world_stacks} Continents broken down based on the distribution of cities into our three major clusters.
	}
\end{figure}

We color our cities according to their major cluster and show them on a world map in Fig.~\ref{fig:world_clusters}. In Fig.~\ref{fig:world_stacks}, we show the breakdown of cities from each continent into the various clusters. We calculate PH for only four major cities in Oceania, so we cannot draw strong conclusions from the cluster assignments of those cities. Among the other regions, we observe that North America has the largest proportion of cities with gridlike street layouts and the smallest proportion of cities with non-gridlike layouts. By contrast, Europe has the smallest proportion of cities with gridlike street layouts. This is consistent with the common perception that North American cities are much more gridlike than European cities. In all regions, we also observe that a large fraction of the cities are interrupted grids. Additionally, we observe that South America, Africa, and Asia have similar distributions of city types.

Interestingly, from the map in Fig.~\ref{fig:world_clusters}, South America, Asia, and Africa appear to have areas that are dominated by specific major clusters. 
We observe non-gridlike cities in the northern part of South America, whereas we see gridlike cities along its east coast. In Africa, most of the non-gridlike cities occur along the western coastline. In Asia, most of the gridlike and non-gridlike cities occur in East Asia, whereas Southeast Asia is dominated by interrupted grids.
Across the map, there appears to be a potential equatorial band of non-gridlike cities. We do not have an explanation for these patterns, but they are fascinating and seem worthy of future research efforts.

\subsubsection{Comparison of Our Classification of Cities to that of Louf and Barthelemy \cite{2014louf}}
\label{sss:existing}


We compare our results to the city classification of Louf and Barthelemy \cite{2014louf}, who associated each city with a conditional probability distribution that captures the area and shape of its blocks. We choose their method as a point of comparison because they studied a wide range of cities and (like us) codified cities from a block-based perspective. They used the word ``fingerprint'' as a monicker for their block-based representation of cities. In our method, we codify cities according to their PHs, which we generate using the level-set construction of Section~\ref{ss:ls}. Both the approach of \cite{2014louf} and our approach capture information that is based on city blocks, although our PH representation differs substantially from the fingerprints of \cite{2014louf}.

Louf and Barthelemy clustered cities into four groups, whereas we have chosen to cluster our cities into three groups. In \cite{2014louf}, European and North American cities largely inhabit the same cluster (group three in \cite{2014louf}), but they appear in distinct subclusters, demonstrating that there is a substantive difference between cities from the two regions. Our method finds that North America has the largest proportion of cities with gridlike streets among all of the regions and that Europe has the smallest proportion of such cities.

In contrast to the above situation, Africa, Asia, and South America have a fairly balanced composition of city types, with a potential equatorial band of non-gridlike cities. Louf and Barthelemy observed several clusters (groups one, two, and four in \cite{2014louf}) that occur predominantly in Africa, Asia and Oceania (which they combined into one entity), and South America. Notably, none of our clusters are as dominant as group three (which they described as having heterogeneous block sizes and shapes) of \cite{2014louf}, although we do observe that our cluster of cities with interrupted grids (such cities are characterized in part by their heterogeneous block sizes) is also our largest cluster.

Now that we have compared our results to those of \cite{2014louf}, we briefly compare and contrast the types of information that the two methods can capture. Recall that our level-set construction for PH generates filtered simplicial complexes that first consist of streets and then expand outward to absorb the blocks between them. The PH of such a filtered simplicial complex thereby gives a low-dimensional representation of the original image of a city street network. Because irregularly shaped blocks are absorbed into the surrounding streets at a different pace than regular blocks, we capture information about the regularity of each block. Louf and Barthelemy's method also uses information about the regularity of block shape. See Eq.~(3.2) in \cite{2014louf} for a precise mathematical statement of how they measured the regularity of blocks. It is related to a subset of so-called ``compactness measures'' \cite{Gillman2002} (which are used in the study of gerrymandering \cite{DBLP:journals/corr/abs-1803-02857, 2018arXiv180805860D}) that compare the area of a shape to the area of a circle in which the shape is circumscribed.

Because the original image of a city street network includes information about the spatial relationships between blocks, the PH that results from our approach also encodes some of this information. By contrast, Louf and Barthelemy's fingerprints do not encode information about the spatial relationships of blocks to each other. Additionally, our method captures information from dead ends, which Louf and Barthelemy discarded.

Overall, although both our approach and that of \cite{2014louf} use a block-based representation to characterize cities, there are subtle differences in the way that the two approaches encode block information. Nevertheless, the commonality of a block-based perspective results in some similarities. For example, the clusters that result from the two approaches seem to be based heavily on block size and regularity. However, our approach appears to prioritize spatial relationships between different clusters of blocks (specifically, whether blocks are arranged in a grid); such information is not captured in the approach of \cite{2014louf}. Consequently, the two approaches capture different city morphologies, and we expect them to be useful as complementary techniques for studying structures in spatial complex systems.


\subsection{Snowflakes} 
\label{ss:snowflakes}

\begin{figure}[htbp!]
	\centering
	\begin{subfigure}[t]{0.15\textwidth}
		\centering
		\includegraphics[width=\textwidth, height=\textwidth, keepaspectratio]{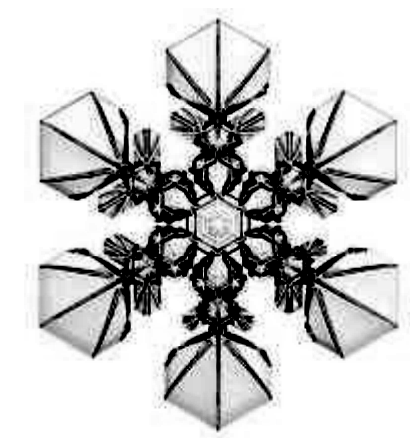}\\
		\includegraphics[width=\textwidth, height=\textwidth, keepaspectratio]{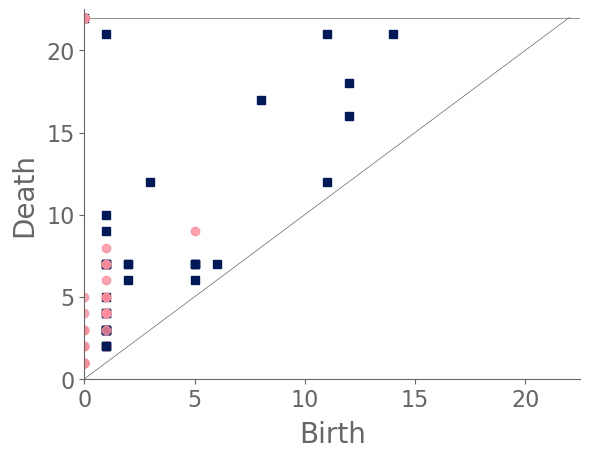}
		\caption{\label{fig:snowflake_a}}
	\end{subfigure}
	\begin{subfigure}[t]{0.15\textwidth}
		\centering
		\includegraphics[width=\textwidth, height=\textwidth, keepaspectratio]{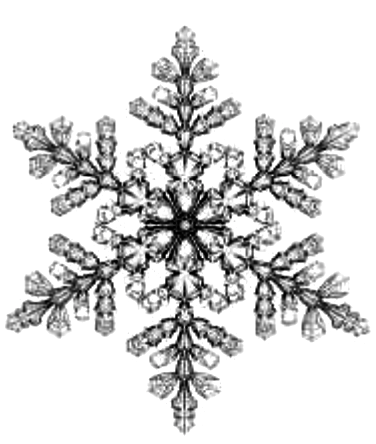}\\
		\includegraphics[width=\textwidth, height=\textwidth, keepaspectratio]{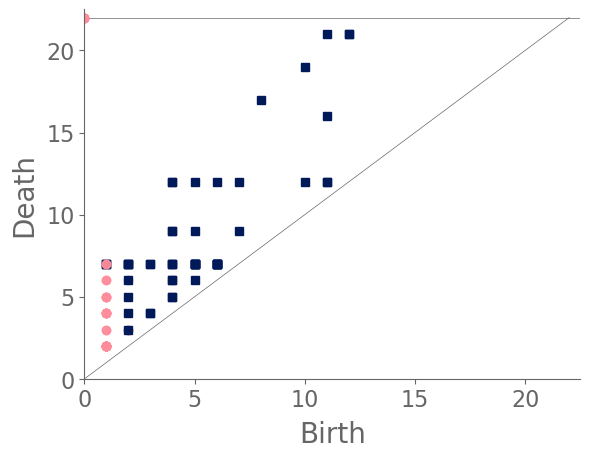}
		\caption{\label{fig:snowflake_b}}
	\end{subfigure}
	\begin{subfigure}[t]{0.15\textwidth}
		\centering
		\includegraphics[width=\textwidth, height=\textwidth, keepaspectratio]{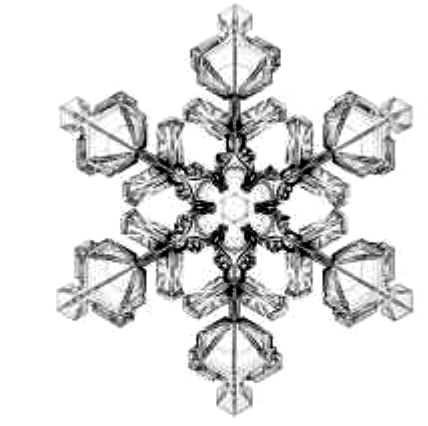}\\
		\includegraphics[width=\textwidth, height=\textwidth, keepaspectratio]{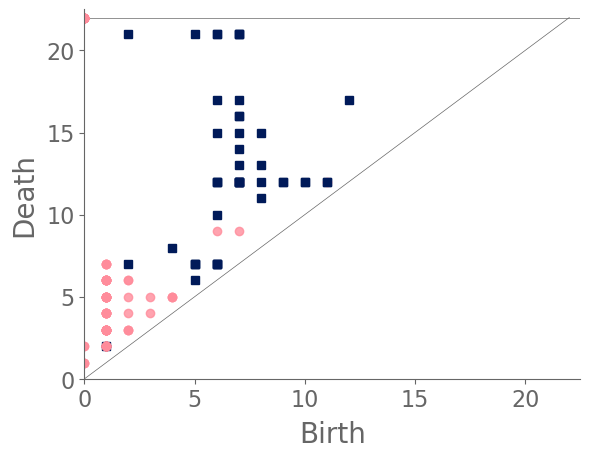}
		\caption{\label{fig:snowflake_l}}
	\end{subfigure}
	\caption{\label{fig:snowflakes} Snowflakes can have a variety of crystalline structures, as we illustrate with (a) Snowflake A, (b) Snowflake B, and (c) Snowflake D. We show the snowflake structures in the top row and their associated PDs in the bottom row. We show the structures of our full set of snowflakes in Fig.~\ref{fig:snowflakes_app}. [The images in the top row are from \cite{2019libbrecht}.]
	}
\end{figure}

\begin{figure}[htbp!]
	\centering
	\includegraphics[width=.45\textwidth]{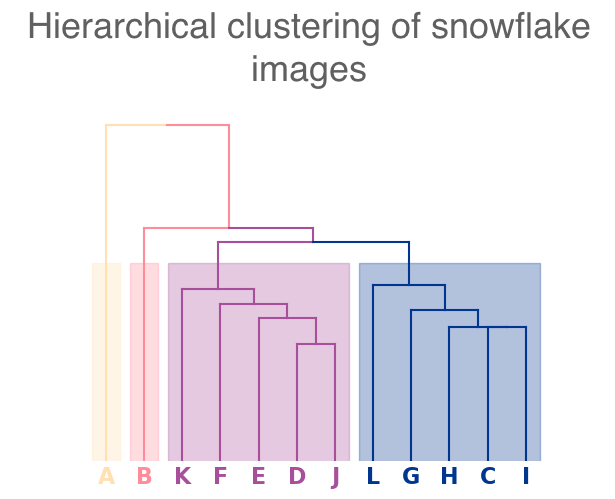}
	\caption{\label{fig:snowflake_dendro} Dendrogram from clustering the snowflakes in Fig.~\ref{fig:snowflakes_app}.
	}
\end{figure}

As a second application that uses empirical data, we consider snowflake crystals \cite{2019libbrecht}. We start with twelve different images (from \cite{2019libbrecht}) of snowflakes with different crystalline structures. (See Fig.~\ref{fig:snowflakes_app} in Appendix~\ref{app:snowflakes}.) Using the GNU Image Manipulation Program \cite{gimp}, we threshold these grayscale images (using a thresholding setting of 205) to create black-and-white images. From the black-and-white images, we compute level-set complexes and PHs, and we then perform average-linkage hierarchical clustering on the PDs to produce the dendrogram in Fig.~\ref{fig:snowflake_dendro}.

The images of snowflakes consist of edges (the black lines in our images) and cells (the white spaces that are bounded by the edges). We refer to the outer areas that extend from the center of the snowflakes as their ``points''. The twelve snowflakes have fairly regular crystalline structures, so our computation of PH predominantly records information about the distribution of cell sizes in a snowflake. The inherent hexagonal nature of snowflakes and the regularity of their crystalline structures largely overwhelms our ability to use PH to glean information about their spatial relationships and irregularities. Constructing a simplicial complex that is better suited to capturing information about images with a large number of regular structures may yield better results.

Examining the clusters (see Fig.~\ref{fig:snowflake_dendro}) reveals that snowflake A and snowflake B each reside in their own cluster, and the remaining snowflakes split into two clusters. Snowflake A's PD [see Fig.~\ref{fig:snowflakes}(a)] is dominated by a feature in $H_1$ with an early birth and a late death. (See the blue square in the top-left corner of the diagram. This arises from the large feature that is formed by the bold ring near the center of the snowflake.
 None of the other snowflakes have a bold central ring. More generally, we observe few features in the PD for snowflake A. By contrast, snowflake B's features are largely close to the diagonal [see Fig.~\ref{fig:snowflakes}(b)] because the initial manifold of the snowflake does not have large holes. Notably, we do not observe any points in the top-left region of its PD. The cell sizes in snowflake B are smaller than those in most of the other snowflakes, and even its central ring structure includes a large number of small holes. The remaining snowflakes either have more large holes than snowflake B, or they do not have a bold central ring like the one in snowflake A. We also note the PD of snowflake B is much closer than that of snowflake A to those of the other snowflakes.


\subsection{Spiderwebs}
\label{ss:spiderwebs}

Our final application is to the topology of spiderwebs. In 1948, Peter Witt began research on the effects of drugs on spiders to test whether garden spiders would shift their web-building hours if they were administered drugs. Witt found that drugs affect the size and shape of the webs that are produced by spiders~\footnote{Interestingly, whiskey itself produces webs \cite{whiskey2019}.}. He also found that higher doses of most drugs (e.g., 100 $\mu$g per spider, as opposed to 10 $\mu$g per spider) tend to lead to larger changes in the shapes of webs, including yielding more irregular webs. 
Witt eventually published more than $100$ papers and several books on the behavior of spiders and spider webs. For more information on his experiments with psychotropic substances and spiders, see his 1971 review article \cite{witt_spiders}. In a 1995 technical briefing \cite{spiders}, NASA (which was inspired by Witt's research) proposed that spiders who were administered more toxic substances produce webs that are more deformed (in comparison to a web that is spun by a drug-free spider) than less toxic substances. Additionally, using techniques from statistical crystallography, they concluded that spiders fail to complete more sides of their webs when they are under the influence of more toxic substances.

\begin{figure}[htbp!] 
	\vspace{.5cm}
	\centering
	\includegraphics[width=.45\textwidth]{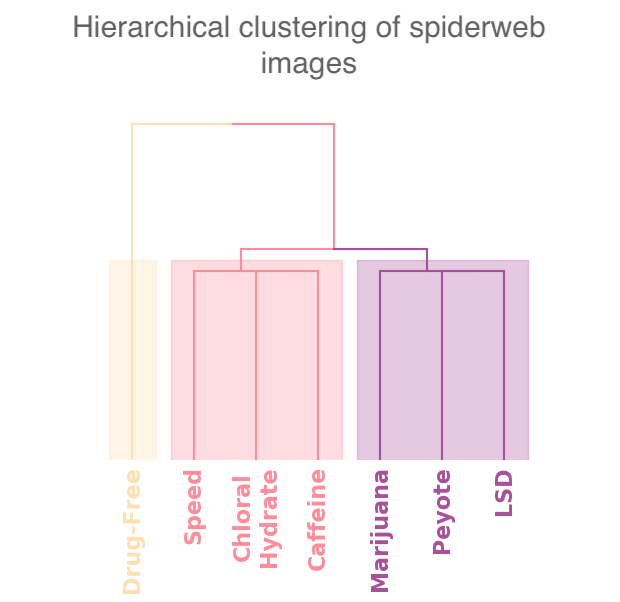}
	\caption{\label{fig:spider_dendro} Classification of webs that were spun by spiders under the influence of various psychotropic  substances. 
	}
\end{figure}

\begin{figure*}[htbp!] 
	\centering
	\begin{subfigure}[t]{0.2\textwidth}
		\centering
		\includegraphics[width=\textwidth, height=\textwidth]{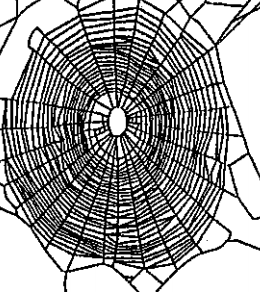}\\
		\includegraphics[width=\textwidth, height=\textwidth]{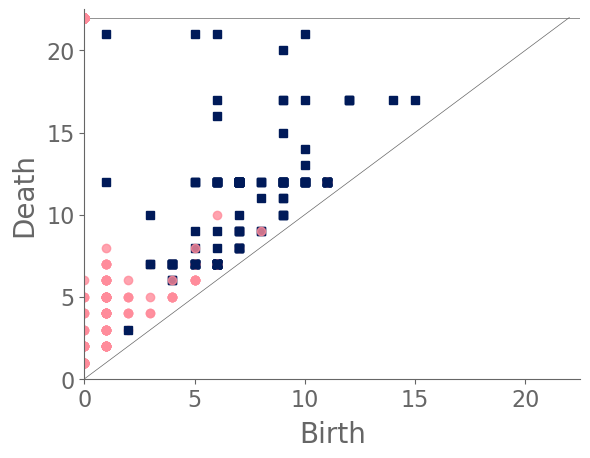}
		\caption{}
	\end{subfigure}
	\begin{subfigure}[t]{0.2\textwidth}
		\centering
		\includegraphics[width=\textwidth, height=\textwidth]{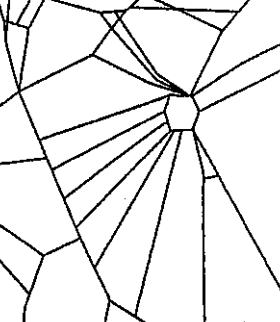}\\
		\includegraphics[width=\textwidth, height=\textwidth]{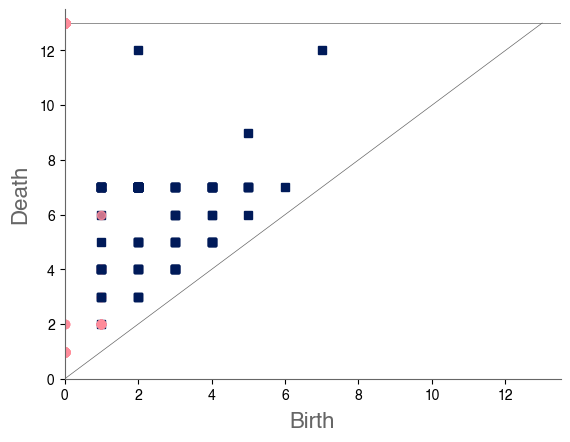}
		\caption{}
	\end{subfigure}
	\begin{subfigure}[t]{0.2\textwidth}
		\centering
		\includegraphics[width=\textwidth, height=\textwidth]{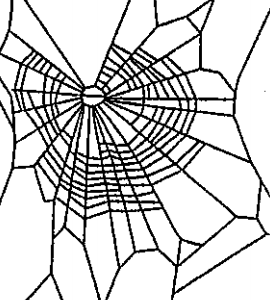}\\
		\includegraphics[width=\textwidth, height=\textwidth]{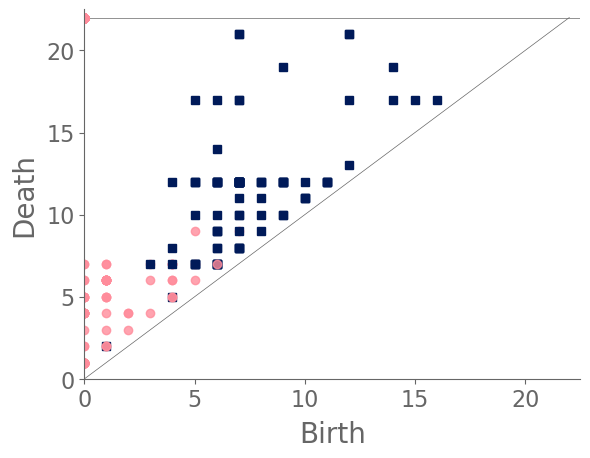}
		\caption{}
	\end{subfigure}
	\begin{subfigure}[t]{0.2\textwidth}
		\centering
		\includegraphics[width=\textwidth, height=\textwidth]{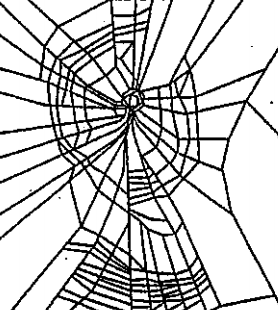}\\
		\includegraphics[width=\textwidth, height=\textwidth]{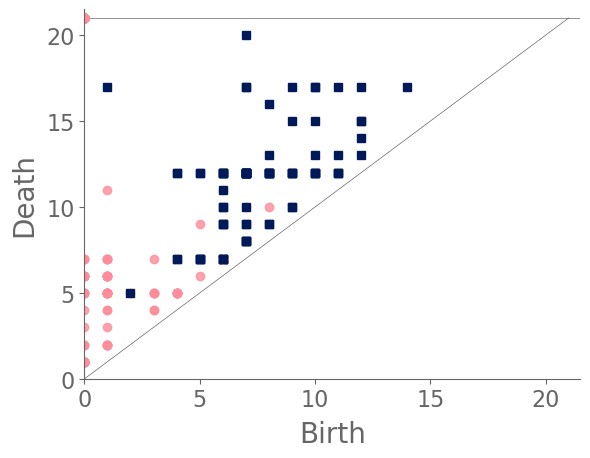}
		\caption{}
	\end{subfigure}\\
	\vspace{.5cm}
	\begin{subfigure}[t]{0.2\textwidth}
		\centering
		\includegraphics[width=\textwidth, height=\textwidth]{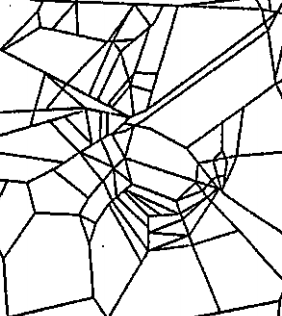}\\
		\includegraphics[width=\textwidth, height=\textwidth]{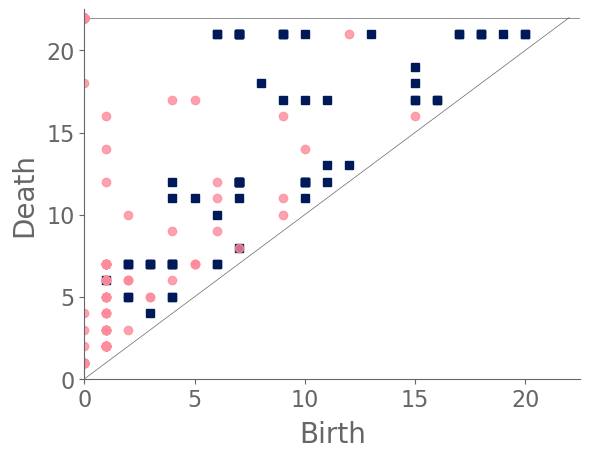}
		\caption{}
	\end{subfigure}
	\begin{subfigure}[t]{0.2\textwidth}
		\centering
		\includegraphics[width=\textwidth, height=\textwidth]{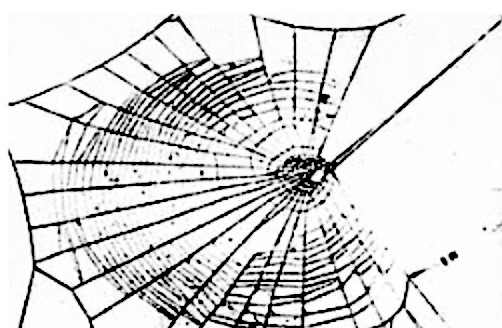}\\
		\includegraphics[width=\textwidth, height=\textwidth]{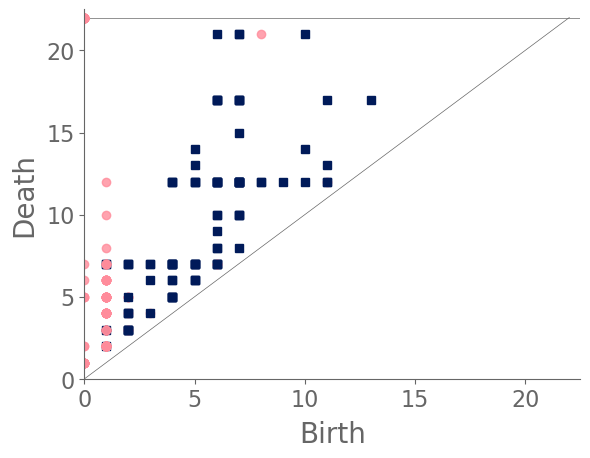}
		\caption{}
	\end{subfigure}
	\begin{subfigure}[t]{0.2\textwidth}
		\centering
		\includegraphics[width=\textwidth, height=\textwidth]{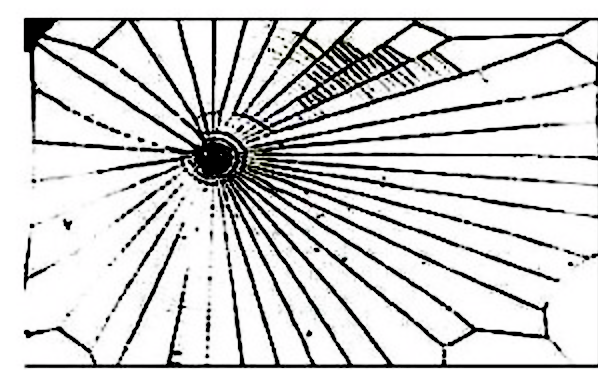}\\
		\includegraphics[width=\textwidth, height=\textwidth]{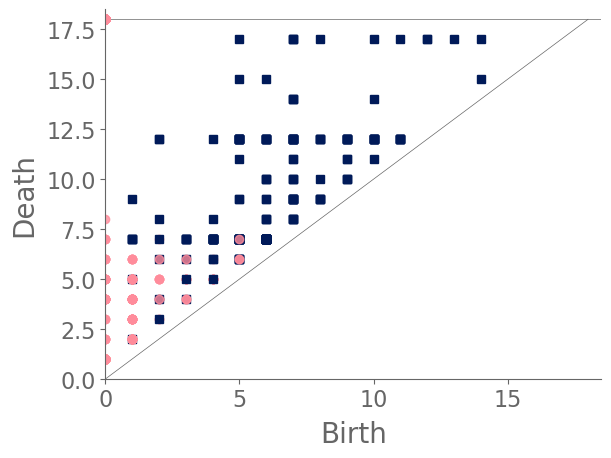}
		\caption{}
	\end{subfigure}
	\caption{\label{fig:spiders_raw} Webs spun by a drug-free spider and spiders that were under the influence of various psychotropic substances, with the associated PD displayed beneath each web. We compare the webs of (a) a drug-free spider with webs spun by spiders that were under the influence of (b) chloral hydrate (which is used in some sleeping pills), (c) marijuana, (d) speed, (e) caffeine, (f) peyote, and (g) LSD. [The images for panels (a)--(e) are from \cite{spiders}, and the images for panels (f) and (g) are from \cite{witt_spiders}.]
	}
\end{figure*}

In our case study of PH in spiderwebs, we use five images from the NASA technical briefing \cite{spiders} and two images from Witt \cite{witt_spiders} of webs that were spun by spiders under the influence of a variety of psychotropic substances, threshold grayscale images to turn them into black-and-white images (using a thresholding setting of 205 in the GNU Image Manipulation Program), apply our level-set construction to compute PH, and perform average-linkage hierarchical clustering to yield the dendrogram in Fig.~\ref{fig:spider_dendro}. We show the images of the spiderwebs and their associated PDs in Fig.~\ref{fig:spiders_raw}. 

Our classification places the drug-free spider into its own cluster. The spiderweb of the drug-free spider is characterized by a clear central hole, threads radiating outward at approximately even intervals, and completed rings of threads that surround the center. We place the webs that were spun by spiders under the influence of marijuana, peyote, and LSD into the same cluster. In these webs, there is a clearly identifiable center; and most of the radial threads are evenly-spaced, straight, and radiate outward directly from the center. However, for the webs in this cluster, rings of threads are either difficult to see or are incomplete. The final cluster consists of webs that were spun by spiders under the influence of chloral hydrate, caffeine, and speed. In the caffeinated spider's web, one cannot even clearly identify a center \footnote{The web that was produced by the caffeinated spider is always fun to point out when giving presentations.}. One can locate a center in the webs of the spiders that were under the influence of speed or chloral hydrate (a sedative that is used in sleeping pills), but many of the radial threads do not join the center and some of the radial threads are not straight. Almost no complete rings of thread are visible in any of the three webs in this cluster.

\section{Conclusions} \label{conc}

It is important to exploit spatial information in the study of spatial complex systems. In this paper, by using new methods of computing persistent homology that take spatial information into account, we presented several applications of topological data analysis to spatial networks. We showed that topological methods are capable of characterizing network structures and detecting structural differences from images of various spatial networks. We also demonstrated, using both synthetic examples and networks from empirical data, that such methods are able to provide insights into large-scale network structures that complement those from traditional techniques of network analysis. As an extended case study, we examined the morphology of street networks in cities, and we used spatial TDA to compare and contrast (1) different regions of the same city and (2) different cities. We hope that our examples help illustrate some ways in which topological methods, especially ones that directly incorporate spatial information in their formulation, can be useful for the analysis of spatial complex systems.


\appendix

\section{\label{app:snowflakes}Additional Snowflake Images}

\begin{figure*}[htbp]
	\centering
	\begin{subfigure}[t]{0.15\textwidth}
		\centering
		\includegraphics[width=\textwidth, height=\textwidth]{figures/snowflakes/raw/1}
		\caption{\label{afig:snowflake_a}}
	\end{subfigure}
	\begin{subfigure}[t]{0.15\textwidth}
		\centering
		\includegraphics[width=\textwidth, height=\textwidth]{figures/snowflakes/raw/2}
		\caption{\label{afig:snowflake_b}}
	\end{subfigure}
	\begin{subfigure}[t]{0.15\textwidth}
		\centering
		\includegraphics[width=\textwidth, height=\textwidth]{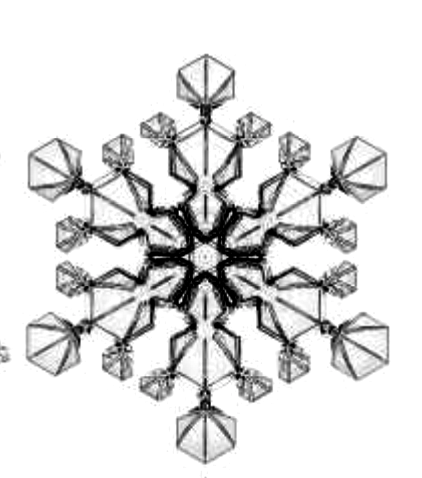}
		\caption{\label{afig:snowflake_c}}
	\end{subfigure}
	\begin{subfigure}[t]{0.15\textwidth}
		\centering
		\includegraphics[width=\textwidth, height=\textwidth]{figures/snowflakes/raw/4}
		\caption{\label{afig:snowflake_d}}
	\end{subfigure}
	\begin{subfigure}[t]{0.15\textwidth}
		\centering
		\includegraphics[width=\textwidth, height=\textwidth]{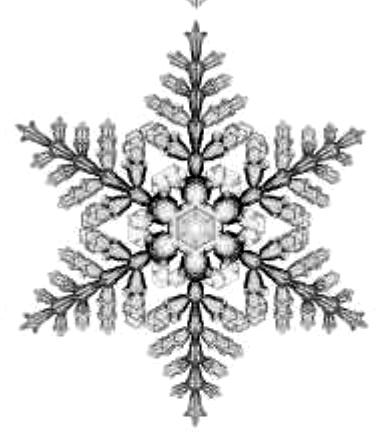}
		\caption{\label{afig:snowflake_e}}
	\end{subfigure}
	\begin{subfigure}[t]{0.15\textwidth}
		\centering
		\includegraphics[width=\textwidth, height=\textwidth]{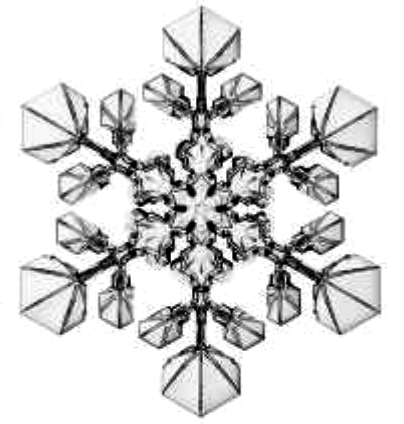}
		\caption{\label{afig:snowflake_f}}
	\end{subfigure}\\
	\begin{subfigure}[t]{0.15\textwidth}
		\centering
		\includegraphics[width=\textwidth, height=\textwidth]{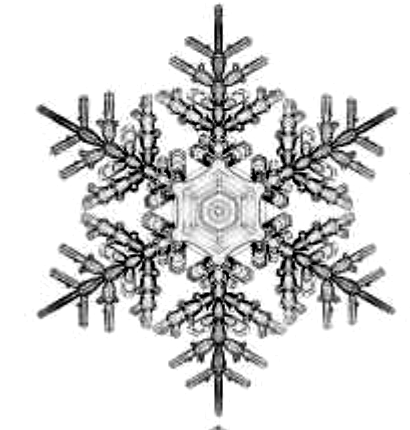}
		\caption{\label{afig:snowflake_g}}
	\end{subfigure}
	\begin{subfigure}[t]{0.15\textwidth}
		\centering
		\includegraphics[width=\textwidth, height=\textwidth]{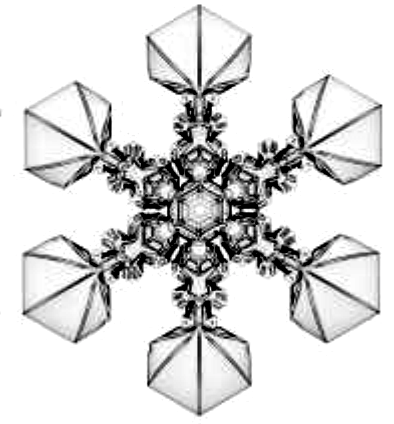}
		\caption{\label{afig:snowflake_h}}
	\end{subfigure}
	\begin{subfigure}[t]{0.15\textwidth}
		\centering
		\includegraphics[width=\textwidth, height=\textwidth]{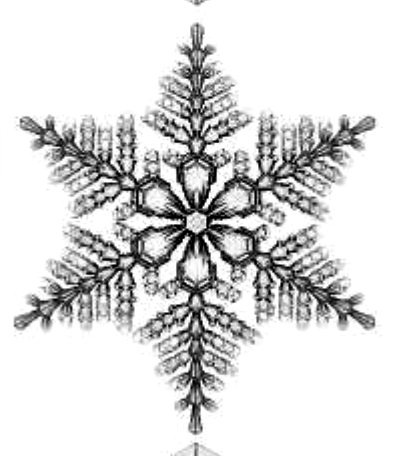}
		\caption{\label{afig:snowflake_i}}
	\end{subfigure}
	\begin{subfigure}[t]{0.15\textwidth}
		\centering
		\includegraphics[width=\textwidth, height=\textwidth]{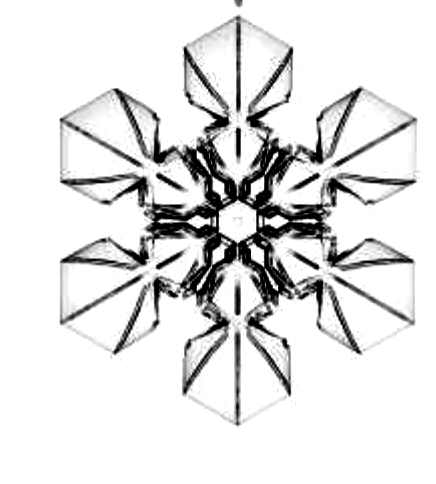}
		\caption{\label{afig:snowflake_j}}
	\end{subfigure}
	\begin{subfigure}[t]{0.15\textwidth}
		\centering
		\includegraphics[width=\textwidth, height=\textwidth]{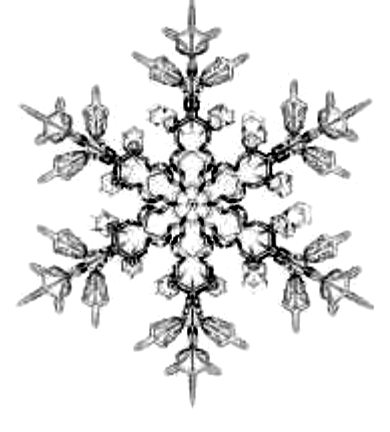}
		\caption{\label{afig:snowflake_k}}
	\end{subfigure}
	\begin{subfigure}[t]{0.15\textwidth}
		\centering
		\includegraphics[width=\textwidth, height=\textwidth]{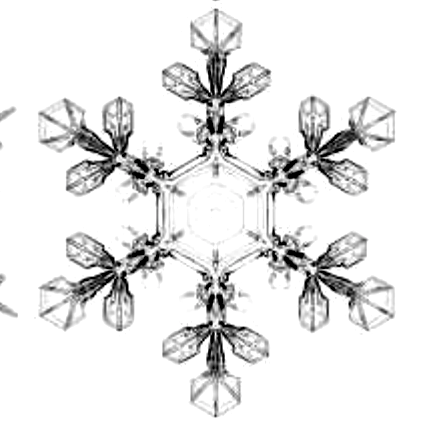}
		\caption{\label{afig:snowflake_l}}
	\end{subfigure}
	\caption{\label{fig:snowflakes_app} The full set of twelve snowflake images that we examined in Section~\ref{ss:snowflakes}. We label these snowflakes using the panel labels from this figure. We show Snowflake A in panel (a), Snowflake B in panel (b), and so on. [These images are from \cite{2019libbrecht}.]
	}
\end{figure*}

In Fig.~\ref{fig:snowflakes_app}, we show the images of all twelve snowflakes that we examined.

\begin{acknowledgments}

We thank Marc Barthelemy, Heather Zinn Brooks, Hanbaek Lyu, Elizabeth Munch, Stan Osher, Nina Otter, Giovanni Petri, Bernadette Stolz, and an anonymous referee for helpful comments. We are particularly grateful to Joshua Gensler for his many helpful comments on both our paper and our code. We also acknowledge support from the National Science Foundation (grant number 1922952) through the Algorithms for Threat Detection (ATD) program.

\end{acknowledgments}




%


\end{document}